\documentclass[11pt]{article} 
\usepackage{hyperref} 
\usepackage{graphicx}

\begin{document} 
 
\title{The Tibetan Singing Bowl} 
 
\author{D. Terwagne$^\dagger$ and J.W.M. Bush$^\ddagger$\\
\\\vspace{6pt} $^\dagger$GRASP, D\'epartement de Physique, \\\vspace{6pt} 
Universit\'e de Li\`ege, B-4000 Li\`ege, Belgium\\\vspace{6pt} 
$^\ddagger$Department of Mathematics,\\\vspace{6pt} 
Massachusetts Institute of Technology, 02139 Cambridge, USA } 
 
\maketitle

\begin{abstract} 
The Tibetan singing bowl is a type of standing bell. Originating from Himalayan fire cults as early as the 5th century BC, they  have since been used in religious ceremonies, for shamanic journeying, exorcism, meditation and shakra adjustment.  A singing bowl is played by striking or rubbing its rim with a wooden or leather-wrapped mallet. The sides and rim of the bowl then vibrate to produce a rich sound. When the bowl is filled with water, this excitation can cause crispation of the water surface that can be followed by more complicated surface wave patterns and ultimately the creation of droplets. We here demonstrate the means by which the Tibetan singing bowl can levitate droplets.
This is a sample arXiv article illustrating the use of fluid dynamics videos.
\end{abstract} 


The first part of the video shows a water-filled Tibetan bowl rubbed by a leather mallet that excites vibrations via a "stick-slip" process. The rim deflection indicates that 
the main deformation mode is the fundamental one associated with four nodes and four anti-nodes along the rim. We proceed by reporting the form of the flow induced by the moving rim, specifically, the evolution of the free surface with increasing rim forcing.

The bowl is completely filled with water, and its fundamental deformation mode excited acoustically. A loudspeaker produces a sinusoidal sound at a frequency $f_0=188 \,\rm{Hz}$ that corresponds to the fundamental frequency excited by rubbing the rim.

\begin{figure}[h!]
\begin{center}
\includegraphics[width=12cm]{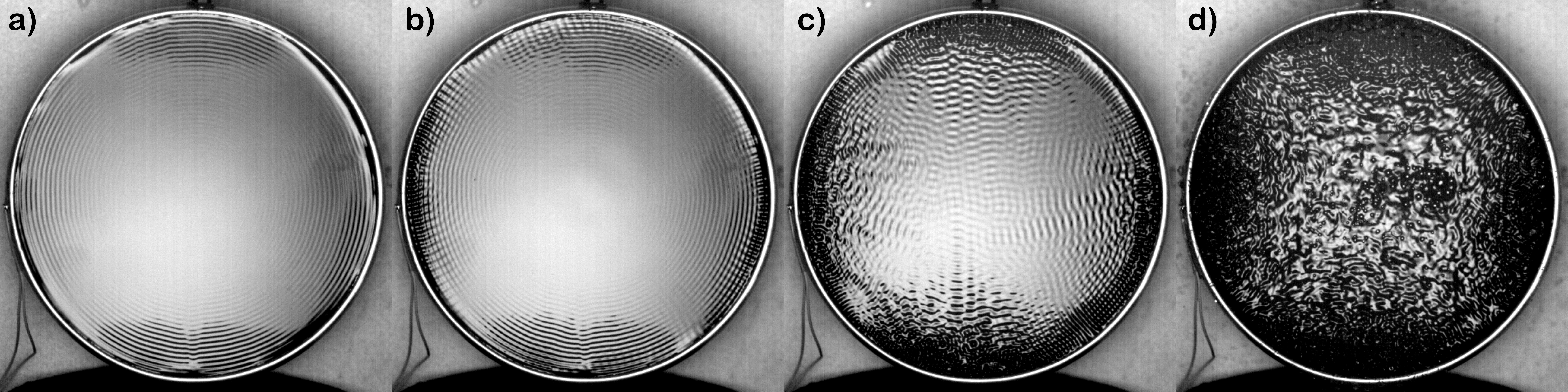}
\end{center}
\caption{Evolution of the surface waves in a water-filled bowl excited with a frequency of $f_0=188\,\rm{Hz}$. The amplitude $A$ of deformation at anti-nodes is increasing from left to right : a) $A=13\,\mu\rm{m}$, b) $A=20\,\mu\rm{m}$, c) $A=44\,\mu\rm{m}$, d) $A=115\,\mu\rm{m}$. \label{T1surf}}
\end{figure}

The vibration of the water surface is forced by the horizontal oscillation of the rim. When the amplitude of the rim oscillation is small, axisymmetric capillary waves with frequency commensurate with the excitation frequency appear on the liquid surface. Though almost invisible to the naked eye, they can be readily detected by appropriate lighting of the liquid surface. In Fig. \ref{T1surf}.a, one can see the evolution of these axial waves, as viewed from above, with increasing forcing.

When the rim deflection amplitude is further increased, circumferential waves appear at the water's edge (see Fig. \ref{T1surf}.b). The amplitude of these waves grows rapidly, and is larger than that of the axial waves; moreover, their frequency is half that of the axial waves (the excitation frequency $f_0$). The circumferential waves correspond to classic edge-induced Faraday waves \cite{Faraday1831} (see Fig. \ref{T1surf}.b). More complicated wave modes appear at higher excitation amplitude (see Fig. \ref{T1surf}.c).

At sufficiently high excitation amplitude, water droplets are ejected from the edge of the vessel (see Fig. \ref{T1surf}.d), specifically from anti-nodes of the wall oscillations. The ejected droplets jump, bounce, slide and roll on the water surface until eventually coalescing.

With a more viscous fluid (e.g. silicone oil of viscosity 10 cSt) the waves are less pronounced, and the liquid surface may be induced to oscillate up and down at a small distance from the wall's anti-node. When a droplet of the same liquid is deposited on the surface, it may bounce, levitated by the underlying wave field. The air film between the drop and the liquid surface is squeezed and regenerated at each successive bounce, its sustenance precluding coalescence \cite{Couder2005, Gilet2008} and enabling droplet levitation in the Tibetan singing bowl.


%
\end{document}